\pdfoutput=1
\documentclass[a4paper,showpacs,preprintnumbers,amsmath,amssymb,pre]{revtex4}
\usepackage[utf8]{inputenc}
\usepackage{amsopn}
\usepackage{bm,bbm}
\usepackage{amsmath}
\usepackage{amssymb}
\usepackage{multirow}
\usepackage{hyperref}
\usepackage{graphicx}
\usepackage{subfig}
\newcommand{\ket}[1]{\ensuremath{|#1\rangle}}

\newcommand{\HH}{\mathcal{H}}

\newcommand{\ee}{\mathrm{e}}
\newcommand{\ii}{\mathrm{i}}

\newcommand{\1}{{\rm 1\hspace{-0.9mm}l}}
\newcommand{\Id}{\1}
\usepackage{todonotes}

\begin{document}

\title{Cooperative quantum Parrondo's games}

\author{{\L}ukasz Pawela}
\email{lukasz.pawela@gmail.com}
\affiliation{Institute of Theoretical and Applied Informatics, Polish Academy
of Sciences, Ba{\l}tycka 5, 44-100 Gliwice, Poland}

\author{Jan S{\l}adkowski}
\email{jan.sladkowski@us.edu.pl}
\affiliation{Institute of Physics, University of Silesia,
Uniwersytecka 4, 40-007 Katowice, Poland}

\begin{abstract}
Coordination and cooperation are among the most important issues of game 
theory. Recently, the attention turned to game theory on graphs and social 
networks. Encouraged by interesting results obtained in quantum evolutionary 
game analysis, we study cooperative Parrondo's games in a quantum setup. The 
game is modeled using multidimensional quantum random
walks with biased coins. We use the GHZ and W  entangled states as the initial 
state of the coins. Our analysis shows than an apparent paradox in
cooperative quantum games and some interesting phenomena can be observed.
\end{abstract}

\date{28/II/2013}
\pacs{03.67.-a, 02.50.Le, 05.40.Fb}

\maketitle
\section{Introduction}

Game theory is a branch of mathematics that formalizes competitions 
with rational rules and rational players
\cite{osborne1994course}. This theory has  broad applications in a great 
number of fields, from biology to social
sciences and economics. Recently, much attention has been focused on 
transferring concepts of game theory to the
quantum realm. Of course, quantum games are games in the standard sense but 
the approach allows for quantum phenomena in the course of the game 
\cite{piotrowski_invitation_2002, multiqubit_entangling}. Some classical game 
theoretical issues can be extended to allow quantum strategies. Usually, the 
set of quantum strategies is much larger than a ``classical'' one and 
entanglement implies more complex behavior of agents than the ``classical 
mixing'' of strategies \cite{osborne1994course} in such games. An $N$-player 
quantum game can be defined as a 4-tuple
\begin{equation}
	\Gamma = (\HH,\rho,\mathcal{S},\mathcal{P}),
\end{equation}
where $\HH$ is a Hilbert space, $\rho$ is a quantum state (i.e. a density
matrix), $\mathcal{S} = \{S_i\}_{i=1}^N$ is the set of possible player's
strategies and $\mathcal{P} = \{P_i\}_{i=1}^N$ is a set of payoff functions for
the players. A quantum strategy $s_i^\alpha \in S_i$ is a completely positive
trace preserving (CPTP) map. The payoff function of $i$-th player $P_i$ assigns
to a given set of player's strategies $\{s_j^{\alpha_j}\}_{j=1}^N$ a real 
number
-- the payoff. Usually, the set of strategies is limited to unitary operators
and the payoff is determined via a measurement of an appropriate variable.
Access to  such rich strategy sets allows for some spectacular results. For
example, it has been shown that if only one player is aware of the quantum
nature of the system, he/she will never lose in some types of games
\cite{eisert1999quantum}. Recently, it has been demonstrated that a player can
cheat by appending additional qubits to the quantum system
\cite{miszczak2011qubit}. Moreover, one can study the impact of random
strategies on the course of the game~\cite{kosik_quantum_2007}.

The  seminal works of Axelrod~\cite{axelrod1984evolution} and Nowak and 
May~\cite{nowak1992evolutionary} incited the  researchers to investigate the
population structures with local interactions that model   various real social
structures with sometimes astonishing  accuracy. In that  way, evolutionary 
game theory has been married with network structure  analysis. In particular,
the issues of coordination and cooperation with the  involved dilemmas and
efficiency problems have been analysed from  this point   of view
\cite{tomassini2007}. Game theoretical models, although often  unrealistic if
applied to complex human behaviour, provide a simple way of   understanding 
some important aspects  of complex human decisions. Quantum game  theory
approach extends such analyses in an interesting way  \cite{Busemeyer_2006,
abbot, paw_slad, sladkowski2003giffen, edward2003quantum}.  Parrondo's 
paradox, showing that in some cases combination of apparently losing games can 
result in successes,  spurred us on to the analysis of  Parrondo's paradox in 
this context presented  in the present work.

This paper is organized as follows. In Section~\ref{sec:parrondo} we give a
brief description of Parrondo's games, concentrating on the cooperative game. 
In Section~\ref{sec:model} we present our model used for simulation. In
Section~\ref{sec:results} we present results obtained from simulation. Finally,
in Section~\ref{sec:conc} we draw the final conclusions.
\section{Parrondo's games}\label{sec:parrondo}

\subsection{Original paradox}
The Parrondo's paradox \cite{parrondo1996eec} was originally discovered in the 
following context. Consider two coin tossing games, $A$ and $B$.
Let the first
game be a toss of a biased coin with winning probability $p = \frac12 - \epsilon$. The second game is based on two biased
coins and the choice of the coin depends on the current state (pay-off) of the game. Coin $B_1$ is selected if the capital of
the player is a multiple of 3. This coin has a probability of winning $p_1$. 
Otherwise, coin $B_2$ with winning
probability $p_2$ is chosen. Each winning results in a gain of one unit of 
capital, while each loss results in a loss
of one unit of capital. Choosing for example:
\begin{equation}
	p_1 = \frac{1}{10} - \epsilon, \; p_2 = \frac34 - \epsilon,
\end{equation}
results in a losing game $B$. This happens because, the coin $B_1$ is played 
more often than $\frac13$ of the time.
However, if games $A$ and $B$ are interwoven in the described way, the probability of selecting the coin $B_1$ approaches
$\frac13$ thus resulting in a winning game. Furthermore, the capital gain from 
this game can overcome the small capital loss
resulting from game $A$. This construction can be generalized to 
history-dependent games instead of capital-dependent 
ones~\cite{parrondo_new_2000}.

Since its discovery the Parrondo's paradox has been used to describe  
situations where losing strategies can combine to win. There exists deep 
connections between the paradox and a variety of physical phenomena. The 
original Parrondo's games can be described as discrete-time and  
discrete-space flashing Brownian ratchets. This fact has been established  
using discretization of the Fokker-Planck equation. In the recent years, many 
examples from physics to population genetics have been reported in the 
literature, showing the generality of the paradox. Generally, the paradox can 
occur in the case of nonlinear interactions of random behavior with an 
asymmetry. In our case, the nonlinearity is due to switching of the games A  
and B. The asymmetry comes from biased coins. A large number of effects, where 
randomness plays a constructive role,  including but not limited to  stochastic
resonance, volatility pumping, the Brazil nut paradox, can be viewed as being 
in the class of Parrondian  phenomena. For a review of the Parrondo's  paradox
see~\cite{abbott2010asymmetry}. For material regarding modeling Parrondo's 
paradox as a quantum walk see~\cite{meyer2002parrondo,bulger2008position}.
  \subsection{Cooperative Parrondo's games}
  Cooperative Parrondo's games were introduced by Toral
\cite{toral_cooperative_2001}. The scheme is as follows. Consider an ensemble 
of $N$ players, each with his/hers own capital $C_i(t)$, $i = 1, 2, \ldots, N$.
As in the original paradox, we consider two games, A and B. Player $i$ can play
either game A or B according to some rules. The main difference from the
original paradox is that probabilities of winning game B depend on the state 
of players
$i-1$ and $i+1$. For simplicity, we only consider the case when the
probabilities of winning at time $t$, depend only on the present state of the
neighbors, hence the probabilities are given by:
\begin{itemize}
	\item $p_1$ if player $i-1$ is a winner and player $i+1$ is a winner
	\item $p_2$ if player $i-1$ is a winner and player $i+1$ is a loser
	\item $p_3$ if player $i-1$ is a loser and player $i+1$ is a winner
	\item $p_4$ if player $i-1$ is a loser and player $i+1$ is a loser
\end{itemize}
The game, by definition, is a winning one, when the average value of the capital
\begin{equation}
\langle C(t) \rangle = \frac1N \sum_{i=1}^N C_i(t),\label{eq:avpay}
\end{equation}
increases with time. If each agent starts the game with a given capital,  
$C_0$, we define the average capital gain as:
\begin{equation}
	\langle C_{\rm G}(t)\rangle = \frac1N\sum_{i=1}^N (C_i(t) - C_0).
\end{equation}
\section{The model}\label{sec:model}
\subsection{Preliminaries}
There are several known approaches to quantization of Parrondo's
games~\cite{flitney_quantum_2002,gawron_quantum_2005}.
We model a cooperative quantum Parrondo's game as a multidimensional quantum
random walk (QRW)~\cite{flitney_quantum_2004}. The average position of the walker along each axis
determines each player's payoff. As in the classical case, we consider two games, $A$ and $B$. The first game has a
probability of winning $p_0$, while the second has four probabilities 
associated $\{p_i\}_{i=1}^4$. Similar to the
classical case the probabilities of winning game $B$ depend on the state of the neighboring players. The following two
possible schemes of alternating between games $A$ and $B$ are considered
\begin{enumerate}
	\item random alternation, denoted $A+B$
	\item games played in succession $AABBAABB\ldots$, denoted $[2,2]$.
\end{enumerate}

The Hilbert space
associated with the walker
consists of two components: the coin's Hilbert space and the position Hilbert space
\begin{equation}
	\HH = \HH_{\rm c} \otimes \HH_{\rm pos}.
\end{equation}

We introduce two base states in the single coin Hilbert space, the $\ket{L}$ and $\ket{R}$ states. These states
represent the classical coin's heads and tails respectively.

We focus our attention on the three dimensional case (i.e. a three-player 
game). This allows us to limit the size of the quantum system under 
consideration and allows us to handle it numerically. We
assume the state of the walker as
\begin{equation}
	\ket{\Psi} = \ket{C} \otimes \ket{\psi},
\end{equation}
where $C$ is the state of all coins and $\psi$ represents the position of the
walker in a two dimensional space. Furthermore, the position component of the state of the walker $\ket{\Psi}$,
$\ket{\psi}$, is itself a two component system $\ket{\psi} = \ket{\psi_x}\otimes\ket{\psi_y}$. The Hilbert space
$\HH_c$ is a three-qubit space, hence its dimension is $\textnormal{dim}(\HH_c)=8$.

The evolution of the state $\Psi$ is governed by the operator
\begin{equation}
	U = U_{\rm pos}U_{\rm c3}U_{\rm c2}U_{\rm c1},
\end{equation}
where $U_{pos}$ is the position update operator. The position update is based
on the current state of the coins of all players, and the operator is given by
\begin{equation}
	U_{\rm pos} = \sum_{ (A,B,C) \in \atop \{ P_r, P_l \}^{\times 3}} A 
	\otimes B \otimes C \otimes f(A) \otimes f(B)
	\otimes
	f(C),
\end{equation}
where
\begin{equation}
	f(X) = \left\{
	\begin{array}{ccc}
	S & \mathrm{if} & X\equiv P_{\rm r} \\
	S^\dagger & \mathrm{if} & X\equiv P_{\rm l} \\
	\end{array}\right.
\end{equation}
and $S$ is the shift operator in the position space, $S\ket{x} = \ket{x+1}$, 
$P_{\rm r}$ and $P_{\rm l}$ are the
projection operators on the coin states $\ket{R}$ and $\ket{L}$ respectively.
The tossing of the first player's coin when game $A$ is played is given by the operator
\begin{equation}
	U_c = U_0 \otimes \1_{\rm c} \otimes \1_{\rm c} \otimes \1_{\rm pos}
\end{equation}
where $\1_{\rm pos}$ is an identity operator on the entire position space and 
$\1_c$ is an identity operator on a single
coin space. In the case of game $B$, the tossing of the first player's coin is 
realized by the operator
\begin{equation}
	\begin{split}
		U_c &=  U_1 \otimes P_{\rm r} \otimes P_{\rm r} \otimes \Id_{\rm pos} 
		+ U_2 
		\otimes P_{\rm r} \otimes P_{\rm l} \otimes
		\Id_{\rm pos} +\\ &+ U_3 \otimes P_{\rm l} \otimes P_{\rm r} \otimes 
		\Id_{\rm pos} + U_4 \otimes P_{\rm l}
		\otimes P_{\rm l} \otimes \Id_{\rm pos},
	\end{split}
\end{equation}
where $U_k$ are the operators of tossing a single
coin, given by
\begin{equation}
	U_k = \left(
		\begin{array}{cc}
			\sqrt{\rho_k} & \sqrt{1 - \rho_k}\ee^{\ii\theta_k} \\
			\sqrt{1 - \rho_k}\ee^{\ii\phi_k} & -\sqrt{\rho_k}\ee^{\ii(\theta_k +
			\phi_k)}
		\end{array}
	\right),\label{eq:unitary}
\end{equation}
where $k \in \{0,1,2,3,4\}$, $1-\rho$ is the classical probability that the
coin changes its state, and $\phi_k$ and $\theta_k$ are phase angles. The
classical probabilities $p_i$ and the quantum counterparts $\rho_1$
parameterize the Parrondo phenomena in both situations. In general, there is no
numerical relations between $p_i$ and $\rho _i$. Therefore we use different
symbols to avoid misunderstanding. If not stated otherwise we assume the phase
angles to be $\phi_k =\theta_k = \pi/2$ for all $k$. However, in the last
paragraph of Section~\ref{sec:results} we show the influence of the phase
angles on the behavior of the game.
\subsection{Studied cases}
We assume the probabilities $\rho_k$ to be: $\rho_0 = 0.5$, $\rho_1 = \rho_2 = \rho_3 = 0.5$ and study the impact of
the variation of parameter $\rho_4$ on the behavior of the
game. The following special cases of the initial state of the coins  are assumed:
\begin{enumerate}
	\item GHZ state, $\ket{C} = \frac{1}{\sqrt{2}}(\ket{LLL}+\ket{RRR})$
	\item W state, $\ket{C} = \frac{1}{\sqrt{3}}(\ket{LLR}+\ket{LRL}+\ket{RLL})$
	\item separable state, $\ket{C} = \frac{1}{2\sqrt{2}}(\ket{L} -
	\ket{R})^{\otimes 3}$
	\item A semi-entangled state, $\ket{C} =J\ket{LLL}$
\end{enumerate}
In the last case, the operator $J$ is given by~\cite{abbot}
\begin{equation}
	J(\omega) = \exp(\ii \frac{\omega}{2}\sigma_x^{\otimes 3}) = \Id^{\otimes 
	3}\cos\frac{\omega}{2} + \ii \sigma_x^{\otimes 
	3}\sin\frac{\omega}{2}\label{eq:J},
\end{equation}
where $\omega \in [0, \pi/2]$ is a measure of entanglement. In the case of 
$\omega = \frac{\pi}{2}$, the resulting maximally entangled state is of the 
GHZ class:
\begin{equation}
	J\left(\frac{\pi}{2}\right)\ket{LLL} = \frac{1}{\sqrt{2}}\left( \ket{LLL} 
	+ \ii \ket{RRR} \right).
\end{equation}
We investigate the 
following scenarios of games:
\begin{enumerate}
	\item Game A only, denoted $A$
	\item Game B only, denoted $B$
	\item Game A and B chosen randomly, denoted $A+B$
	\item Game A and B played in the sequence: two games of type A, followed by two games of type B, leading to
	AABBAABBAABB\ldots, denoted $[2,2]$
\end{enumerate}

\section{Results and discussion}\label{sec:results}

\begin{figure}[!h]
		\subfloat[separable
		state]{\label{fig:coin_normal}\includegraphics{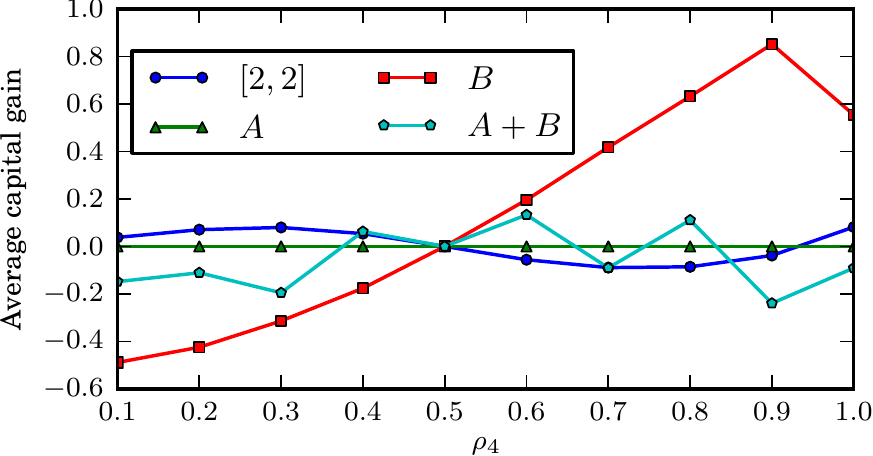}}~
		\subfloat[GHZ state]{\label{fig:coin_GHZ}\includegraphics{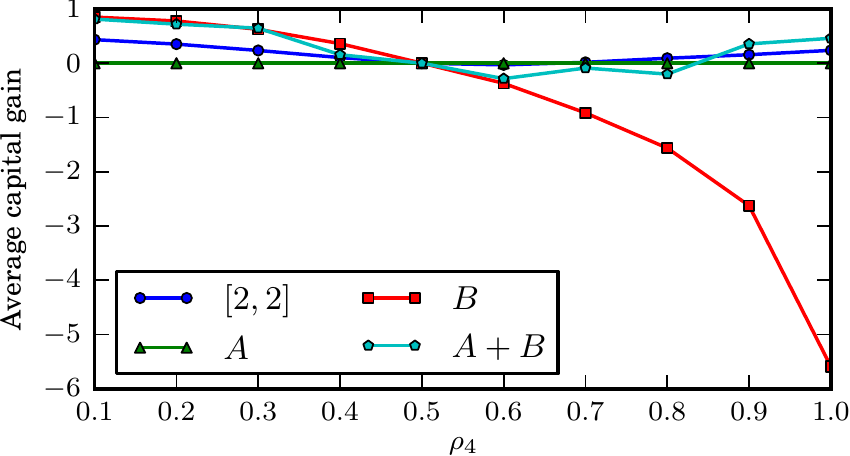}}\\
		\subfloat[W state]{\label{fig:coin_W}\includegraphics{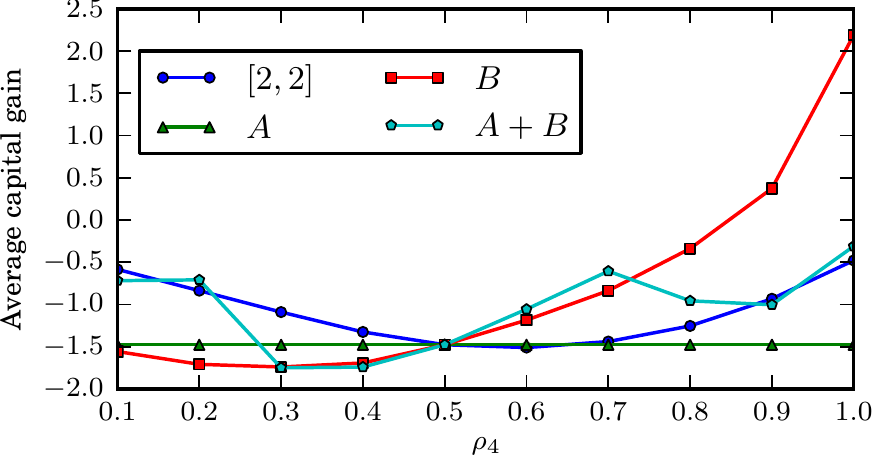}}
	\caption{Average capital gains of all players for different initial states of the coin after 16 rounds of the
	game. Lines are eye-guides.}\label{fig:rhos}
\end{figure}

Figure~\ref{fig:rhos} shows the average capital gains of all players as
defined by Eq.~\eqref{eq:avpay}.
Figures~\ref{fig:coin_normal}, \ref{fig:coin_GHZ} and \ref{fig:coin_W} show results when the initial state of the coin
is separable, the GHZ state and the W state respectively. The capital gains
are taken after 16 rounds of the game. In
each round each players plays exactly once.

\begin{figure}[!h]
	\centering\includegraphics{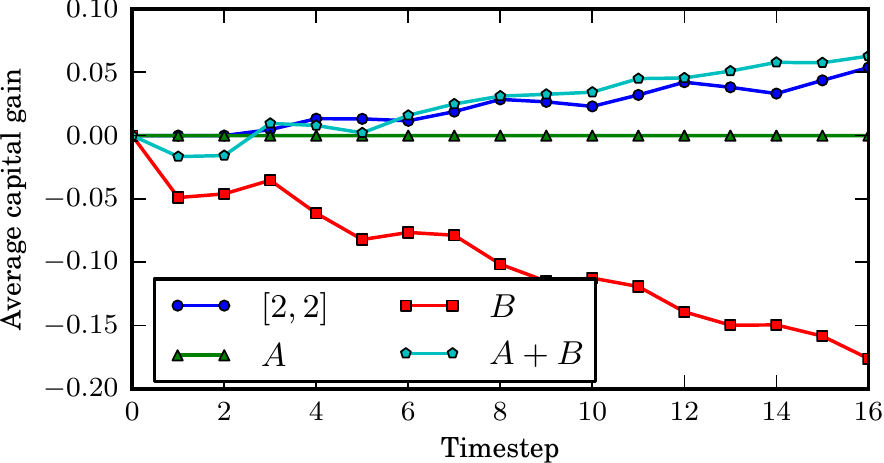}
	\caption{Average capital gains of all players in the case of separable
	initial state, $\rho_4 = 0.4$. Lines are
	eye-guides.}\label{fig:normal_det}
\end{figure}

In the case of a separable initial state, the Parrondo Paradox occurs if
$\rho_4\in[0.1, 0.5)$. Game $[2,2]$
exhibits the Paradox in the whole interval, whereas game $A+B$ is a Parrondo
game only for $\rho_4 = 0.4$. Detailed
results for $\rho_4 = 0.4$ are shown in Figure~\ref{fig:normal_det}. Interestingly, when game B becomes winning, game
$[2,2]$ can become a losing game. This happens for $\rho_4\in(0.5,0.9]$.

\begin{figure}[!h]
	\subfloat[$\rho_4 = 0.7$]{\label{fig:GHZ_comp07}\includegraphics{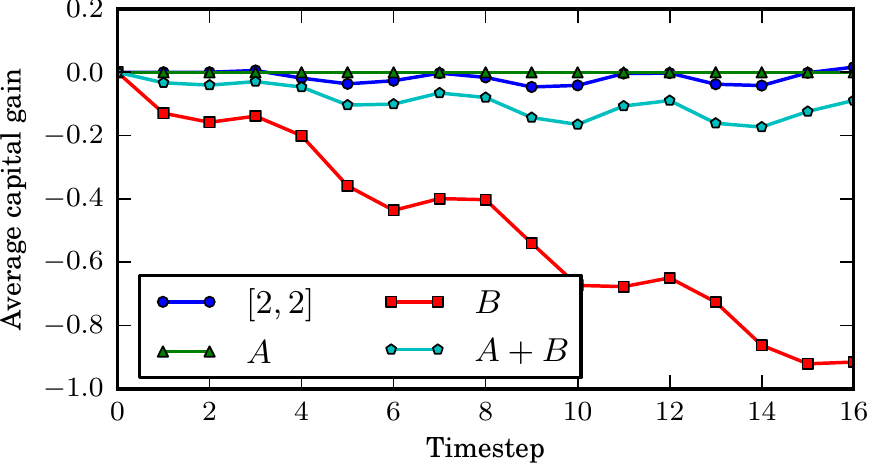}}~
	\subfloat[$\rho_4 = 0.9$]{\label{fig:GHZ_comp09}\includegraphics{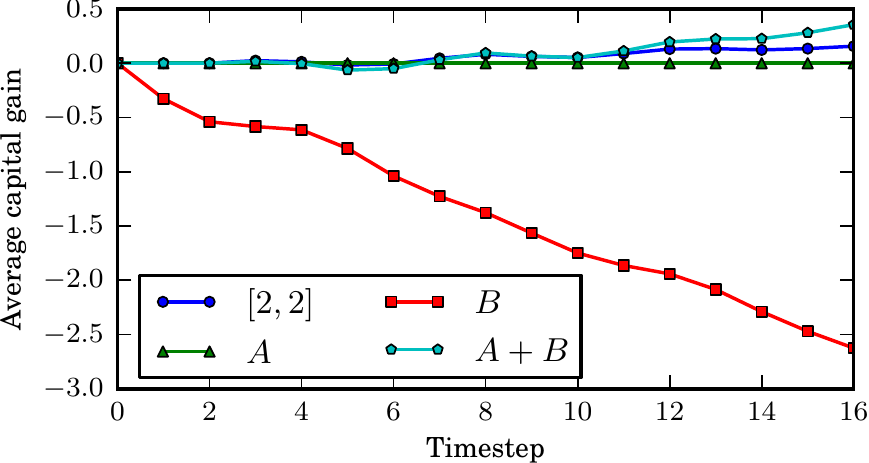}}
	\caption{Comparison of detailed evolutions of capital for the GHZ initial state. Lines are
	eye-guides.}\label{fig:GHZ_comp}
\end{figure}

When the initial state of players' coins is set to be the GHZ state, the nature
of game B changes significantly: the game becomes a winning one for
$\rho_4\in[0.1,0.5)$. As opposed to the previous case, games $[2,2]$ and $A+B$
are also winning in this case. When $\rho_4$ increases further, games $[2,2]$
and $A+B$ become winning games once again, whereas game B becomes a losing 
game. Comparison of detailed evolutions of the average capital gains is shown 
in Figure~\ref{fig:GHZ_comp}. These plots show that, as $\rho_4$ increases, the
behavior of capital changes from oscillatory decreasing (increasing) to linear
decreasing (increasing). Finally, we note that the bigger is the average loss 
of capital in game B, the greater is the capital gain when games $[2,2]$ and
$A+B$ are played.

\begin{figure}[!h]
	\centering\includegraphics{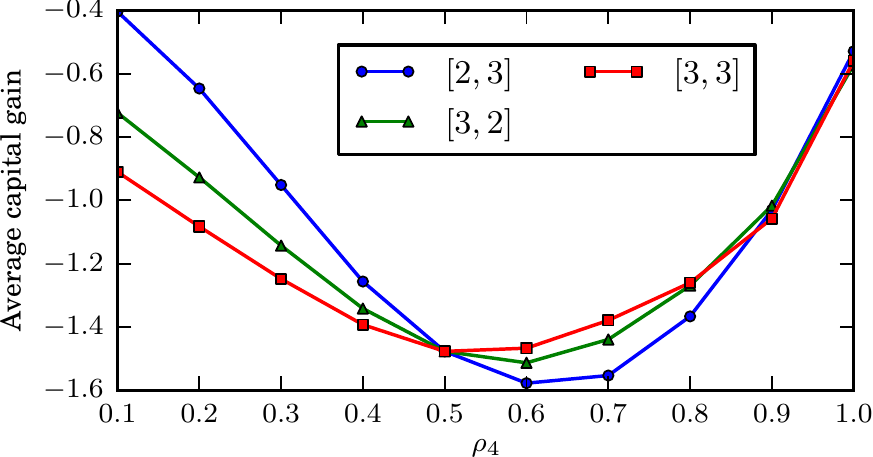}
	\caption{Average capital gains of all players for different games with the 
	W state being the initial state of the
	coins. Lines are eye-guides.}\label{fig:W_games}
\end{figure}

Selecting the W state as the initial one, we find that there is no paradoxical
behavior. This is due to the fact, that for this initial state game A becomes a
losing game as well. To test if this initial state can lead to paradoxical
behavior, we investigated some other game types for this case.
Figure~\ref{fig:W_games} shows the results for games AAABB, AABBB and AAABBB
denoted $[3,2]$, $[2,3]$ and $[3,3]$ respectively. They also do not exhibit any
paradoxical behavior. Therefore, it may be appropriate to propose a method of
distinguishing between the two maximally entangled three qubit states.
Such a possibility might be used i quantum state tomography or initial state 
preparation for some configurations.
\begin{figure}[!h]
	\centering\includegraphics{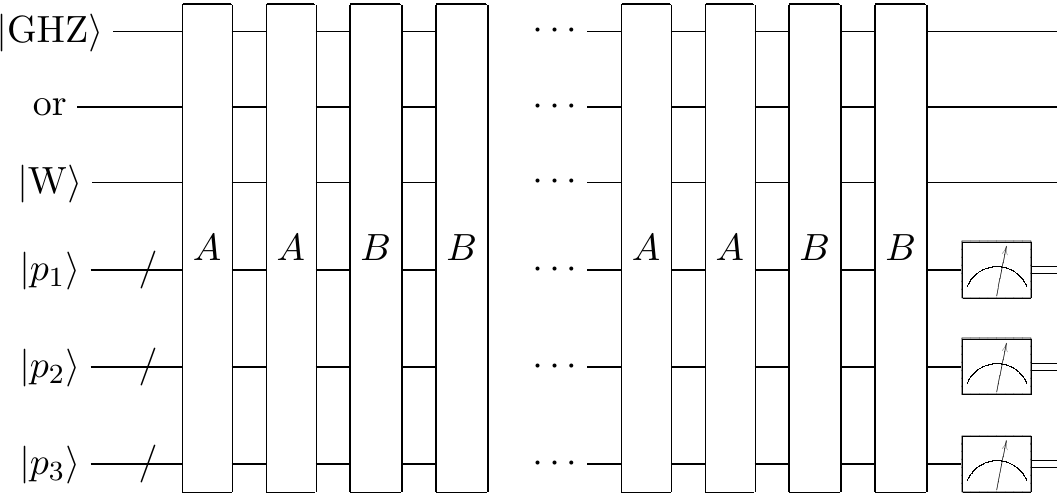}
	\caption{Quantum circuit for distinguishing the W and GHZ 
	states.}\label{fig:circ}
\end{figure}

Consider the quantum circuit depicted in Figure~\ref{fig:circ}. The input 
qubits are the initial state of the coin ($\ket{\mathrm{GHZ}}$ or 
$\ket{\mathrm{W}}$) and registers $\ket{p_i}$ holding the payoff of the $i$-th 
player. After a measurement is performed on these registers, a payoff of each 
player is obtained. Classical addition of these payoffs allows us to  
determine, whether the initial coin state was a GHZ state or a W state.

The change in the behavior of game $A$ when changing from the GHZ to the W 
state can be explained as follows. The fair coin operator acting on the GHZ 
transfers it to the state
\begin{equation}
	\ket{\psi} = \frac14 \left[ (1 - \ii), (\ii - 1), (\ii - 1), (\ii - 1), 
	(\ii - 1), (\ii - 1), (\ii - 1), (1 - \ii)
	\right]^{\rm T}.\label{eq:state_step}
\end{equation}
After another application of the coin flip gate, this state becomes again a
GHZ state. Both, in the GHZ state and the state given by 
Eq~\eqref{eq:state_step} the probabilities of increasing or decreasing a
players payoff are equal. This is not the case for the W state. In this state, 
the ``fair'' coin flip causes the players to lose capital.

\begin{figure}[!h]
		\subfloat[$\omega = 
		0$]{\label{fig:ent_0}\includegraphics{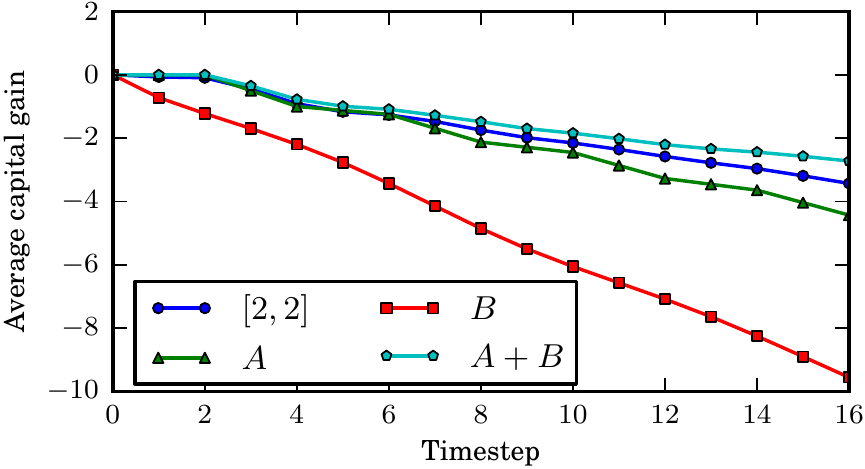}}~
		\subfloat[$\omega = 
		\frac{\pi}{10}$]{\label{fig:ent_1}\includegraphics{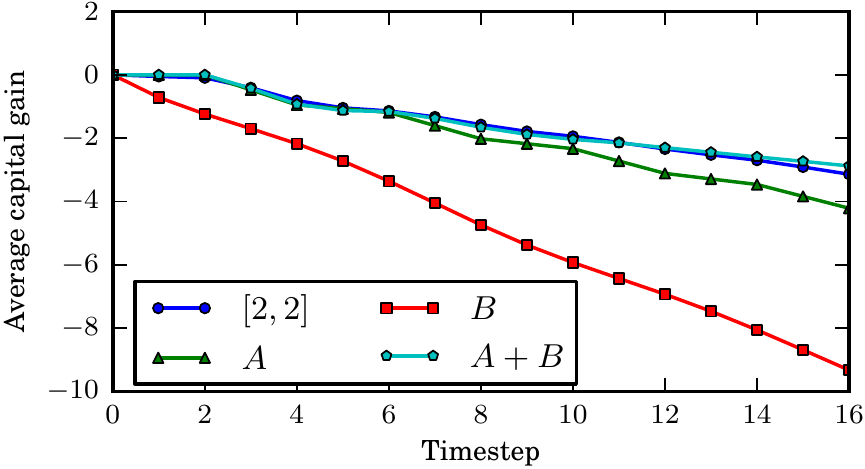}}\\
		
		\subfloat[$\omega = 
		\frac{2\pi}{10}$]{\label{fig:ent_2}\includegraphics{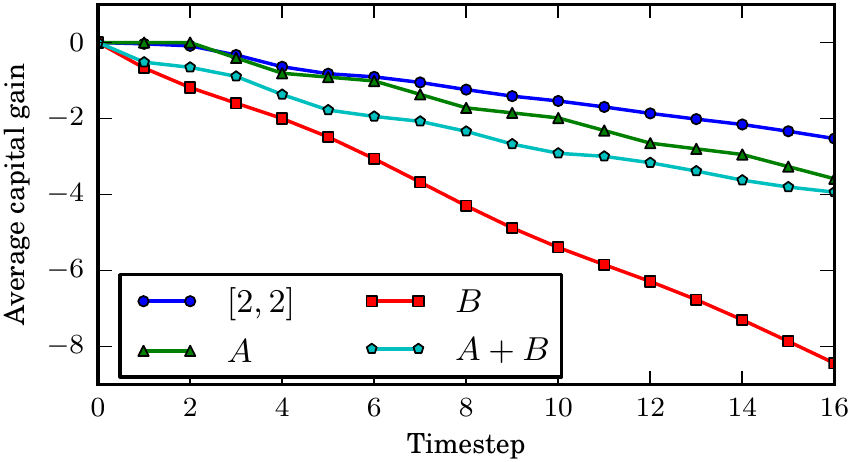}}~
		\subfloat[$\omega = 
		\frac{3\pi}{10}$]{\label{fig:ent_3}\includegraphics{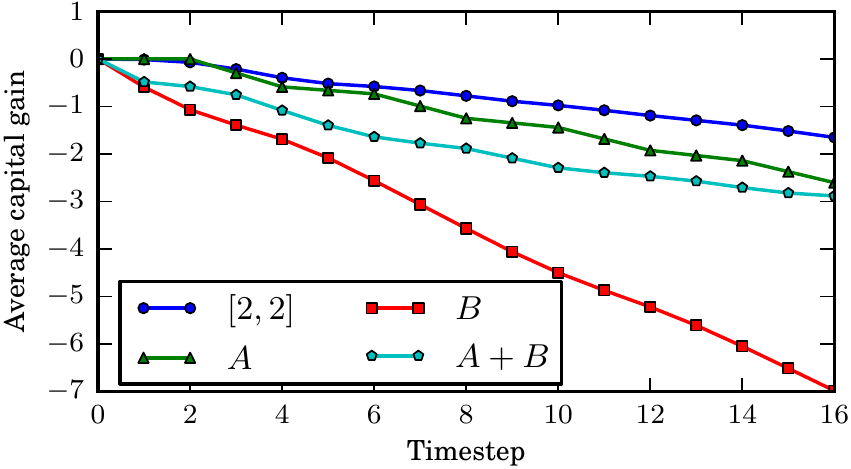}}\\

		\subfloat[$\omega = 
		\frac{4\pi}{10}$]{\label{fig:ent_4}\includegraphics{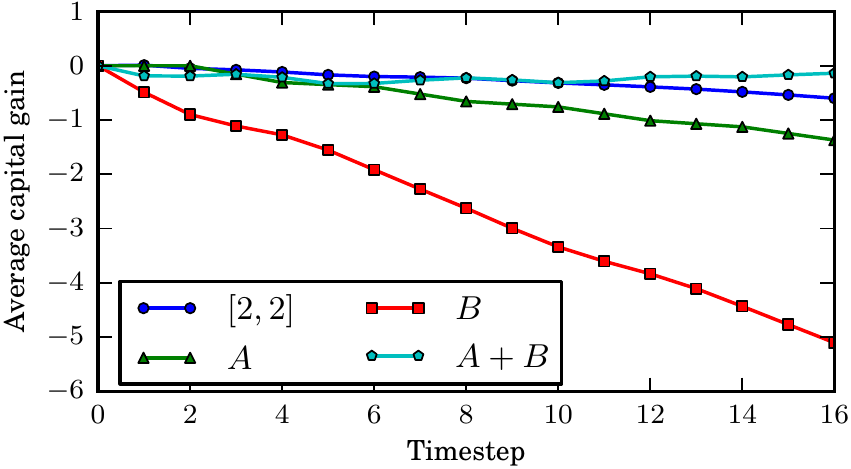}}~
		\subfloat[$\omega = 
		\frac{5\pi}{10}$]{\label{fig:ent_5}\includegraphics{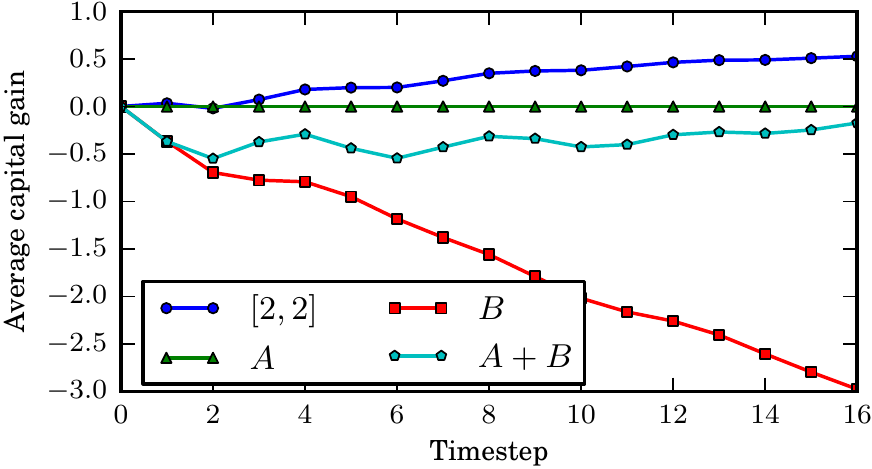}}\\
		
	\caption{Average capital gains of all players for different values of the 
	entanglement $\omega$}\label{fig:ent}
\end{figure}

Figure~\ref{fig:ent} depicts the behavior of the studied games for different 
values of the parameter $\omega$ introduced in Eqn.\eqref{eq:J}. In this 
setup, games A and B are both losing games when $\omega < \frac{\pi}{2}$. 
Furthermore, the games [2,2] and A+B do not exhibit paradoxical behavior. When 
the value of parameter $\omega$ reaches its maximum, two interesting things 
happen: game A becomes a fair game again and, what is more interesting, the 
paradoxical behavior is restored for games [2,2] and A+B.

\begin{figure}
	\centering{\includegraphics{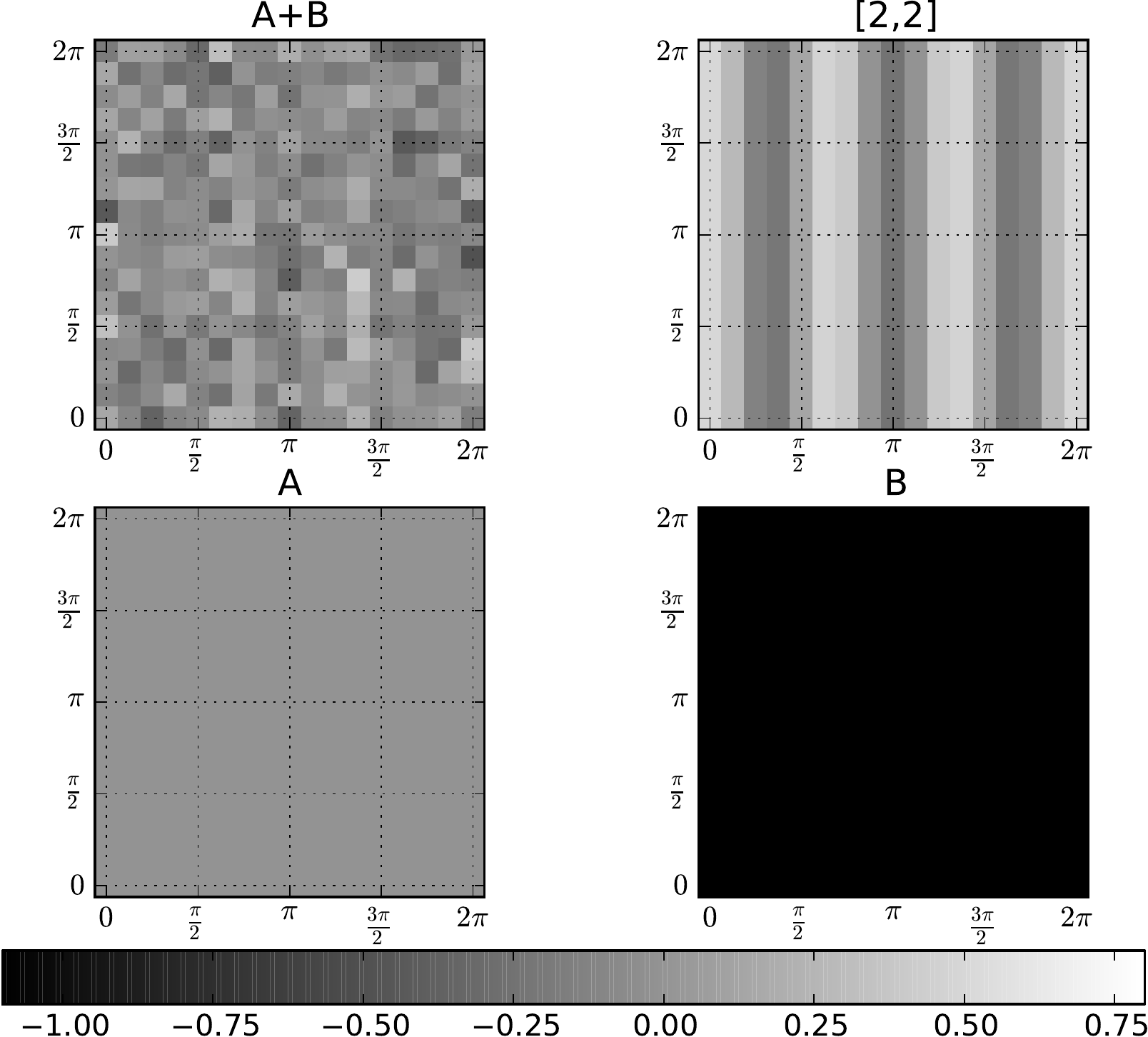}}
	\caption{Map of the average capital gain of all players for the GHZ state 
	for different game setups. The color shows the average capital gain 
	value.}\label{fig:map_GHZ}
\end{figure}

\begin{figure}
	\centering{\includegraphics{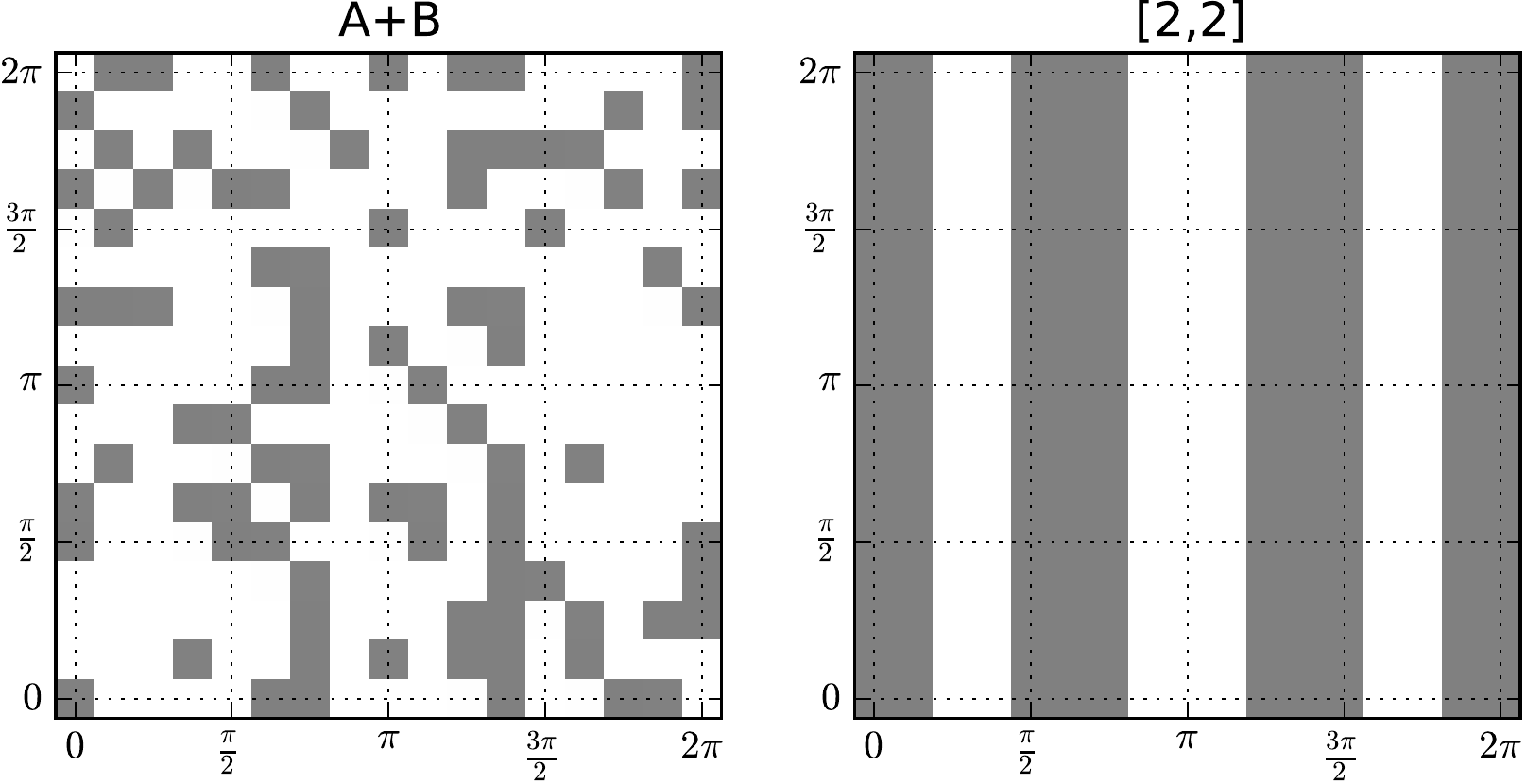}}
	\caption{The gray color marks the values of angles $\phi$ and $\theta$ 
	where the paradox occurs for the GHZ state.}\label{fig:GHZ_possibility}
\end{figure}

\begin{figure}
	\centering{\includegraphics{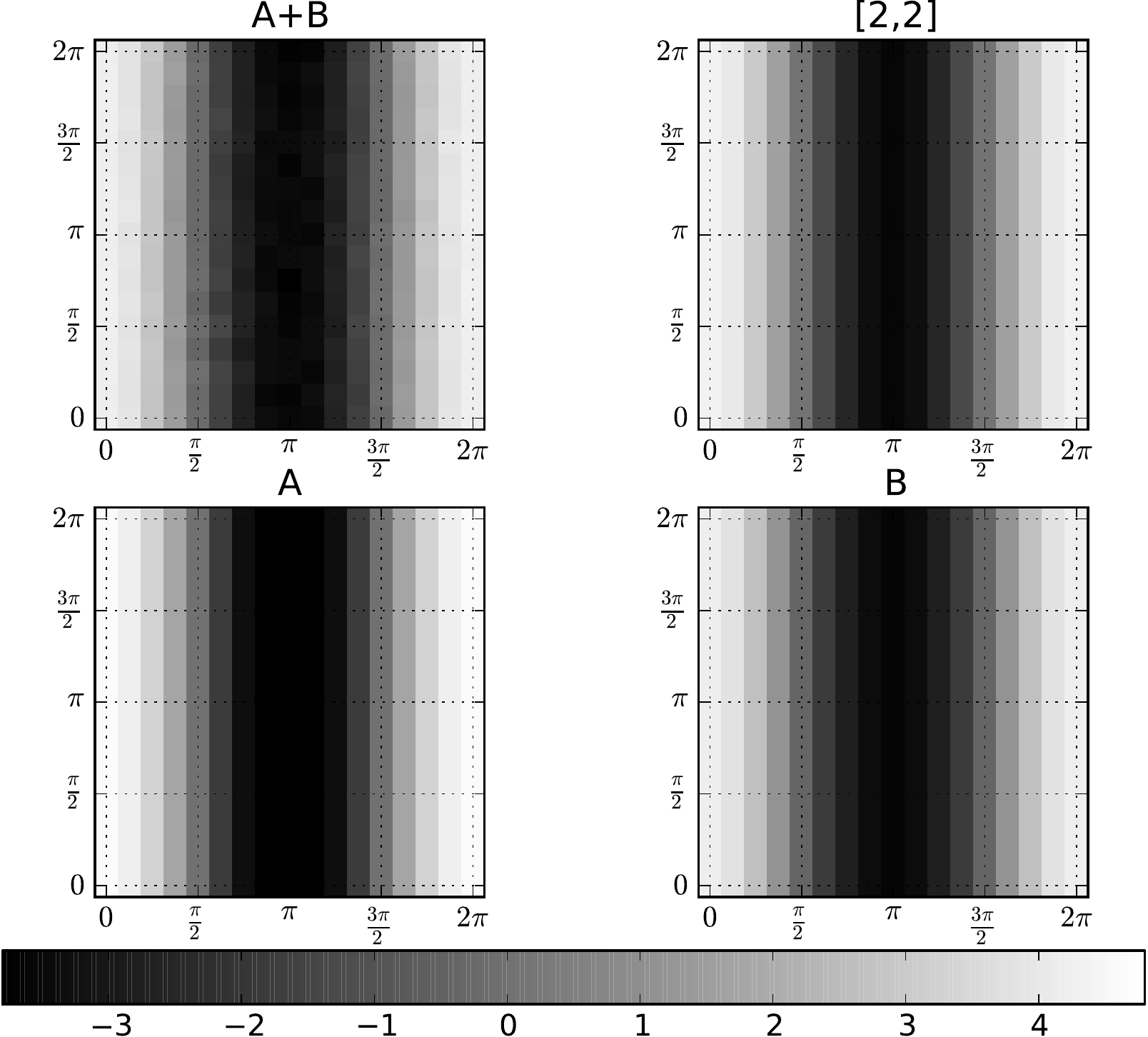}}
	\caption{Map of the average capital gain of all players for the separable 
	state 
	for different game setups. The color shows the average capital gain 
	value.}\label{fig:map_separable}
\end{figure}

\begin{figure}
	\centering{\includegraphics{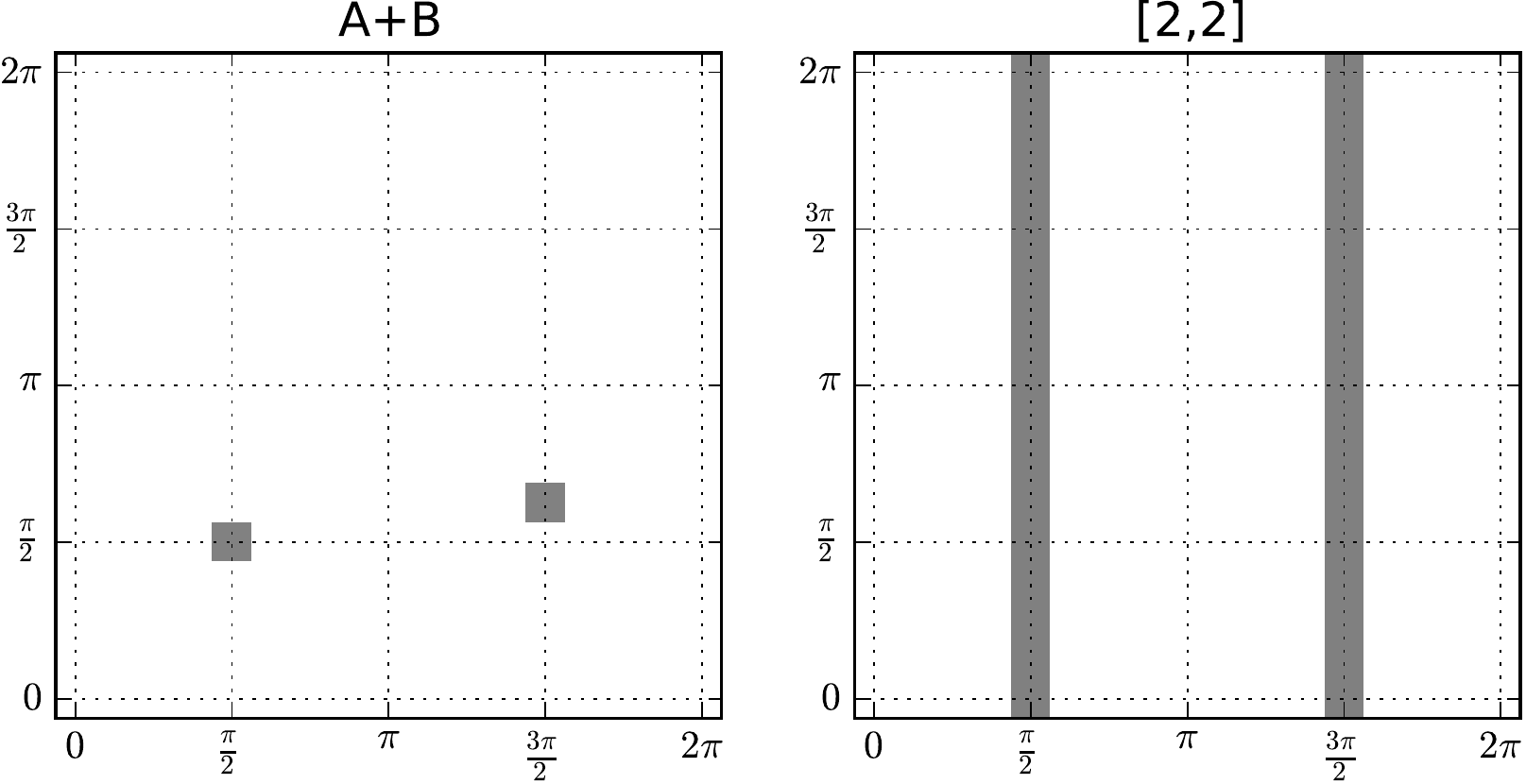}}
	\caption{The gray color marks the values of angles $\phi$ and $\theta$ 
	where the paradox occurs for the separable 
	state.}\label{fig:separable_possibility}
\end{figure}

Finally, we test the impact of the phase angles of the elements of the coin 
operator $\phi$ and $\theta$ defined in Eq~\eqref{eq:unitary}. Maps of the 
average capital gains are shown in Figures~\ref{fig:map_GHZ} and
\ref{fig:map_separable} for the GHZ, separable and W initial coin states 
respectively. In the case of the A+B the results were averaged ten times to 
obtain a smoother picture. The resolution of the plots is $\frac{\pi}{8}$ in 
each direction. Results for the GHZ state show that games A and B 
are insensitive to the phase changes. Game A always remains a fair game and 
game B is always a losing one. The randomness of selection of a specific game 
in the A+B setup has its reflection in the map of the payoffs. The highly 
structured setup od the [2,2] game results in a highly structured map. The 
parameter values for which the paradox occurs are shown in 
Figure~\ref{fig:GHZ_possibility}. Next, we move to the separable state. In 
this case games A and B show a similar structure in the average capital gains. 
This is reflected in games AB and [2,2] for this initial state. 
Figure~\ref{fig:separable_possibility} shows phase angle values for which the 
paradox occurs. Finally, in the case of the W state games A and B are losing 
games and are insensitive to the changes of phase angles. As such, game[2,2] 
is also losing and does not exhibit any change in the average capital gain. 
Game A+B shows some sensitivity to the phase angle values, however it is the 
effect of random switchings between games A and B.

\clearpage
\section{Conclusions}\label{sec:conc}
 We investigated quantum cooperative Parrondo's games modeled using 
multidimensional quantum walks. We studied different initial states of the 
coins of the players: the separable state, the GHZ state and the W state. We
showed that cooperative Parrondo's games can be implemented in the quantum
realm. Furthermore, our analysis shows  how the behavior of a game depends on
the initial state of the coins of all players. One interesting result is that 
if  the initial state of the coins is separable and one game is a winning  one,
then the game where games A and B are interwoven can become a losing game. This
effect does not occur when the initial state of the coins is set to be the GHZ
state. In this case games $A+B$ and $[2,2]$ are always non-losing games. This
shows  that the choice of the initial state may be crucial for the paradoxical 
behaviour. However, the most important result of our work is showing that the
Paradox can also be observed in cooperative quantum games. As a by-product, it
has been shown that the quantum Parrondo paradox may be  used  to easily
distinguish between the GHZ and W states.

\begin{acknowledgements}
Work by  J.~S{\l}adkowski was supported by the Polish National
Science Centre under the project number DEC-2011/01/B/ST6/07197. Work by 
{\L}.~Pawela was supported by the Polish Ministry of Science and Higher 
Education under the project number IP2011 014071. Numerical simulations 
presented in this work were
performed on the ``Leming'' and ``\'Swistak'' computing systems of The Institute of Theoretical and Applied
Informatics, Polish Academy of Sciences.
\end{acknowledgements}

\bibliography{cooperative_parrondo}
\bibliographystyle{apsrev}

\end{document}